# Uncovering a Solvent-Controlled Preferential Growth of Buckminsterfullerene ($C_{60}$) Nanowires

Junfeng Geng,[1,*] Ilia A. Solov'yov,[2,3,*] Wuzong Zhou,[4] Andrey V. Solov'yov[2,3] and Brian F. G. Johnson[1]

[1] *Department of Chemistry, University of Cambridge, Cambridge, CB2, 1EW, United Kingdom*

[2] *Frankfurt Institute for Advanced Studies, Ruth-Moufang-Str. 1, 60438 Frankfurt, Germany*

[3] *A. F. Ioffe Physical-Technical Institute, Politechnicheskaya 26, 194021 St. Petersburg, Russia*

[4] *School of Chemistry, University of St. Andrews, St. Andrews, Fife, KY16, 9ST, United Kingdom*

**The fullerene ($C_{60}$) nanowires, which possess a highly unusual morphology featured by a prism-like central core and three nanobelt-like wings joined along the growth direction to give an overall Y-shaped cross section, have been studied. The experimental observation coupled with theoretical calculation have enabled us to elucidate both the role of the fullerene and of the solvent in the crystallization process, thus opening up an opportunity for the in-depth understanding of the crystal growth mechanism. More generally, the method developed in this work could be extended into understanding the growth of other inorganic nanowires that have both host and guest molecules involved in their crystal lattices.**

---

[*] Corresponding authors. E-mails: jg201@cam.ac.uk; ilia@fias.uni-frankfurt.de.

## 1. Introduction

The growth of one-dimensional (1D) nanocrystals represents an important research topic in crystal engineering for nanotechnology.[1] The growth of 1D fullerene ($C_{60}$) nanocrystals (or nanowires) has proven to be of considerable scientific and technological interest because of the properties associated with the low-dimensionality, quantum confinement effect, and potential electronic, magnetic and photonic applications.[2-6] To this end, there have been a number of reports on the growth, structural characterisation, and application-related investigations of $C_{60}$ nanowires.[7-11] In particular, a recent publication by Miyazawa and co-authors[12] indicates that a pristine $C_{60}$ nanowire, prepared by the liquid-liquid interfacial precipitation method using a pyridine solution of $C_{60}$ and isopropyl alcohol,[13] exhibits electrical conducting behaviour, and the outer $C_{60}$ oxide covering may be potentially used as the dielectric layer for single $C_{60}$ nanowire-based field-effect transistors for nanoelectronics.

In a recent study, we were able to demonstrate that exceptionally long fullerene nanowires, with a length-to-width aspect ratio as large as ~3000, can be grown from 1,2,4-trimethylbenzene (TMB) solution of $C_{60}$.[14] These nanowires, denoted as $C_{60}$·TMB, have been observed to possess a highly unusual shape that retains unchanged even after removing the solvent at elevated temperatures. This excellent property has offered a new approach to the formation of a fullerene-based carbon 1D nanostructure, but importantly, without the involvement of any metal species as the growth catalyst. Consequently, the normally employed post-growth purification process for the removal of metal is no longer necessary, in a marked contrast to the chemical vapour deposition (CVD) technique for growing carbon nanotubes.[14,15]

Broadly speaking, there are two approaches that may be used to prepare 1D nanocrystals in a solution-based synthesis. One is the 'external' method in which surfactant molecules are attached to crystal surface. In this way, the crystal shape may be tuned by varying the growth rates of different crystal facets through preferential surfactant adsorption.[16] This method has been widely employed in colloidal chemistry for making metal nanocrystals of various shapes such as spheres, rods or wires.[17] The second is the 'internal' method in which guest species are introduced into the crystal lattice.[18] In contrast to the first approach, here the crystal shape may be tuned by varying the guest species. This method has been employed in the growth of $C_{60}$ crystals utilising organic solvents as the guest species.[14,18] However, the nature of the $C_{60}$-solvent interactions has so far remained unclear.

The understanding of the exact nature of the $C_{60}$-solvent interaction should provide useful insights into the mechanism of the crystal growth and offer a potential method for the control of the crystal shape and structure. Here we report on our investigations of this important topic from both experimental observations and theoretical calculations. Our approach includes the detailed observation of the crystal morphology and structure using high-resolution electron microscopic techniques, in conjunction with the search of the $C_{60}$-solvent interaction modes in the crystal unit cell, followed by the calculations of the kinetic energies required for the crystal growth along the principal growth axes.

## 2. Experimental Details

Samples of the $C_{60}$·TMB nanowires were prepared by following a previously described method.[4] Scanning electron microscopic (SEM) examination was performed on the LEO-32 electron microscope operated at 5 kV. Samples were directly deposited on a specimen holder (carbon mat) without surface coating of a conducting material. High resolution transmission electron microscopic (HRTEM) images were recorded by a Gatan 794 CCD camera on a JEOL JEM-2011 electron microscope operated at 200 kV. To prepare a specimen for the TEM study, a nanowire sample was first suspended in ethanol, and a drop of the suspension was then deposited on a copper specimen grid coated with holey carbon film. The specimen grid was then placed in a double tilt specimen holder and transferred into the microscopic column. Selected area electron diffractions (SAED) were used in conjunction with HRTEM to determine the crystal structure.

## 3. Theoretical Methods

The interactions between the $C_{60}$ and the TMB molecules in the crystalline lattice were investigated by accounting for the van der Waals and Coulomb interactions between them. The partial charge of each atom in the TMB molecule was calculated within the framework of the *ab initio* density functional theory employing the B3LYP density functional. The calculations have been performed with the Gaussian 03 software package.[19]

The density functional theory (DFT) is based upon a strategy of modeling electron correlation via general functionals of the electron density. Within the DFT one has to solve the Kohn-Sham equations,[20-22] which read as:

$$\left(\frac{\hat{p}^2}{2}+\hat{U}_{ions}+\hat{V}_H+\hat{V}_{xc}\right)\psi_i=\varepsilon_i\psi_i, \qquad (1)$$

where the first term represents the kinetic energy of the *i*-th electron, and $\hat{U}_{ions}$ describes its attraction to the ions in the system, $\hat{V}_H$ is the Hartree part of the interelectronic interaction:

$$\hat{V}_H(\vec{r})=\int\frac{\rho(\vec{r}\,')}{|\vec{r}-\vec{r}\,'|}d\vec{r}\,', \qquad (2)$$

and $\rho(\vec{r}\,')$ is the electron density:

$$\rho(\vec{r}\,')=\sum_{\nu=1}^{N}|\psi_i(\vec{r})|^2, \qquad (3)$$

$\hat{V}_{xc}$ in Eq. (1) is the local exchange-correlation potential, $\psi_i$ are the electronic orbitals and *N* is the number of electrons in the system. The exchange-correlation potential is defined as the functional derivative of the exchange-correlation energy functional:

$$V_{xc}=\frac{\delta E_{xc}[\rho]}{\delta\rho(\vec{r})}, \qquad (4)$$

The approximate functional employed by DFT methods partition the exchange-correlation energy into two parts, referred to as exchange and correlation parts:

$$E_{xc}[\rho]=E_x(\rho)+E_c(\rho). \qquad (5)$$

Physically, these two terms correspond to same-spin and mixed-spin interactions, respectively. Both parts are the functionals of the electron density, which can be of two distinct types: either local functional depending on only the electron density *ρ* or gradient-corrected functional depending on both *ρ* and its gradient, $\nabla\rho$.

In literature, there is a variety of exchange correlation functionals. Below, we refer only to those related to the calculations performed in this work. The local exchange functional is virtually always defined as follows:

$$E_x^{LDA}=-\frac{3}{2}\left(\frac{3}{4\pi}\right)^{1/3}\int\rho^{4/3}d^3\vec{r}. \qquad (6)$$

This form was developed to reproduce the exchange energy of a uniform electron gas. By itself, however, it is not sufficient for the adequate description of a many-body system.

The gradient-corrected exchange functional introduced by Becke[23] based on the local density approximation exchange functional reads as:

$$E_x^{B88} = E_x^{LDA} - \gamma \int \frac{\rho^{4/3} x^2}{1 + 6\gamma \sinh^{-1} x} d^3\vec{r}, \tag{7}$$

where $x = \rho^{-4/3} |\nabla \rho|$ and $\gamma = 0.0042$ is a parameter chosen to fit the known exchange energies of the noble gas atoms.

Analogously to the above exchange functionals, there are local and gradient-corrected correlation functionals, for example, those introduced by Perdew and Wang[24] or by Lee, Yang and Parr.[25] Their explicit expressions are somewhat lengthy and thus we do not present them here but refer to the original papers.

In the pure DFT, an exchange functional usually pairs with a correlation functional. For example, the well-known BLYP functional pairs Becke's gradient-corrected exchange functional (7) with the gradient-corrected correlation functional of Lee, Yang and Parr.[25]

In spite of the success of the pure DFT theory in many cases, one has to admit that the Hartree-Fock theory accounts for the electron exchange the most naturally and precisely. Thus, Becke has suggested[23] functionals which include a mixture of Hartree-Fock and DFT exchange along with DFT correlations, conceptually defining $E_{xc}$ as:

$$E_{xc}^{mix} = c_{HF} E_x^{HF} + c_{DFT} E_{xc}^{DFT}, \tag{8}$$

where $c_{HF}$ and $c_{DFT}$ are constants. Following this idea, a Becke-type three parameter functional can be defined as follows:

$$E_x^{B3LYP} = E_x^{LDA} + c_0 \left( E_x^{HF} - E_x^{LDA} \right) + c_x \left( E_x^{B88} - E_x^{LDA} \right) + E_c^{VWN3} + c_c \left( E_c^{LYP} - E_c^{VWN3} \right). \tag{9}$$

Here, $c_0 = 0.2$, $c_x = 0.72$ and $c_c = 0.81$ are constants, which were derived by fitting to the atomization energies, ionization potentials, proton affinities and first-row atomic energies.[26] $E_x^{LDA}$ and $E_x^{B88}$ are defined in (6) and (7) respectively. $E_x^{HF}$ is the functional corresponding to Hartree-Fock equations. The explicit form for the correlation functional $E_c^{VWN3}$ as well as for gradient-corrected correlation functional of Lee, Yang and Parr, $E_c^{LYP}$, one can find in Ref. *27* and Ref. *25* correspondingly. Note that instead of $E_c^{VWN3}$ and $E_c^{LYP}$ in (9) one can also use the Perdew and Wang correlation functional.[24]

In Gaussian 03, the molecular orbitals, $\psi_i$, are approximated by a linear combination of a pre-defined set of single-electron functions, $\chi_\mu$, known as basis functions. This expansion reads as follows:

$$\psi_i = \sum_{\mu=1}^{N} c_{\mu i} \chi_\mu, \tag{10}$$

where coefficients $c_{\mu i}$ are the molecular orbital expansion coefficients, $N$ is the number of basis functions, which are chosen to be normalized.

The basis functions $\chi_\mu$ are defined as linear combinations of primitive gaussians:

$$\chi_\mu = \sum_p d_{\mu p} g_p ,  \quad (11)$$

where $d_{\mu p}$ are fixed constants within a given basis set, the primitive gaussians, $g_p = g(\alpha, \vec{r})$, are the gaussian-type atomic functions having the following form:

$$g(\alpha, \vec{r}) = c x^n y^m z^l \exp(-\alpha r^2). \quad (12)$$

Here, $c$ is the normalization constant. The choice of the integers $n$, $m$ and $l$ defines the type of the primitive gaussian function: s, p, d or f (for details see Ref. 26). In our calculations we did accounted for all electrons in the system, and employed the standard 6-31G(d) basis set.[26]

The partial charges in the TMB molecule were calculated to fit to the electrostatic potential at points selected according to the Merz-Singh-Kollman scheme (ESP fit method).[28,29] We derived the partial charges of the atoms by fitting the molecular electrostatic potential, because this method has yielded much more promising results[30] than an alternative method based on the Mulliken population analysis.[31] For example, the dipole and quadrupole moments calculated using the charges obtained from the ESP fit method compare favorably to the corresponding experimental gas phase quantities,[30] while the charges obtained using the Muliken population analysis method have been proven to be unsatisfactory.[32]

In the calculations we consider $C_{60}$ and TMB molecules as rigid objects. This assumption is reasonable because the energies involved in the inner dynamics of the molecules are significantly larger than the interaction energies between the molecules.[33,34] Therefore at the temperature of the nanowire growth (~300 K), the constituent molecules remain stable. Freezing the internal degrees of freedom of the molecules significantly reduces the dimensionality of the problem, hence allowing us to study nanowires of larger size. The idea behind the theoretical analysis performed in the present paper is to understand to what extent the intermolecular interactions can explain the large anisotropy of the nanowires.

The energetic of the system was calculated by considering the intermolecular potentials described by several parameters such as the depth of the potential well ($\varepsilon_0$), the bonding length of an atomic pair ($\sigma$), and the atomic charges ($q$), derived from the *ab initio* DFT calculations, as described above. Optimization of the structure of the crystalline unit cell was performed using conjugate gradient method implemented within the MBN Explorer program.[35] The adhesion energies were defined for the system and calculated, in order to determine the energy needed for a unit cell to extend along a specific growth direction. Finally the theoretical results were compared with the experimental observations.

## 4. Results and Discussion

In addition to the nanowires of two wings reported previously,[14] the 1D nanocrystals with three nanobelt-like wings along the growth direction are now observed (Figure 1). The width of these wings vary from one crystal to another. In two extreme cases, this width can be either as large as ~500 nm or as small as almost zero. In the latter case, the nanowires are often curled. These nanowires are typically ~200–600 μm long, and the length-to-width aspect ratio is estimated as large as ~3000. By applying HRTEM and selected area electron diffractions, the crystal structure is determined as orthorhombic, the same as that of the two-wing nanowires.[4]

To understand why the crystal structure is orthorhombic, we first attempted the identification of the shape of the central core along the growth direction in the nanowires. This was successfully done by observing a sample with a broken crystal at its edge parts using the grid tilting mechanism under TEM. Such a shape was identified as being a 1D prism-like, with an overall Y-shaped cross-section (Figure 2a). The angle between any adjacent wings within a crystal is 120°. Since the TMB solvent itself does not crystallize, the initial nucleation of the crystals must first occur from $C_{60}$ molecules. Considering the fact that there are two possible types of nucleation for $C_{60}$ (one is hexagonal close pack (*hcp*), and the other is face centre cubic (*fcc*)),[36] we reasoned in this case that the observed prism-like crystalline core could originate from the hexagonal nucleation of $C_{60}$, but the initial hexagonal structure was significantly distorted in the subsequent growth process because of the inclusion of the TMB solvent molecules (see Figures 2b and 2c for the shape evolution). Consequently, the growth developed towards the formation of an orthorhombic structure.

This view is in agreement with the literature reports that the formation of a hexagonal structure is frequently found when an organic solvent is used for similar crystallizations.[37,38] Also in this work, nanocrystals having a hexagonal shape, as a side-product in the same batch of growth, were observed, which serves as an additional evidence for the above view (Figure 2d). In addition, our view is supported by the observation that there is a high image contrast in TEM around the central core as shown in Figures 1 and 2. The large contrast can be understood in terms of the diffraction contrast caused by the different structural phases between the core and the wings, and at their interface a dark contrast is thereby given.

In the previous report,[14] we proposed a 'central growth' mechanism for the formation of the nanowires but did not have direct experimental evidence found at the time. Here we provide such an evidence to more fully justify our proposal. As can be seen from Figure 3, a tip at the end of the nanocrystals following the growth is observed. Such a short, tail-like tip was extruded at the ends of the nanowires, clearly indicates that the growth of the central core is always slightly pacing ahead of the side wings. In this way, the growth of the central core virtually leads the overall growth of the crystals to develop along the preferential one dimension, whilst the side wings develop in the directions perpendicular to the growth axis.

To understand the growth mechanism in more details, we have considered the van der Waals interactions between $C_{60}$ and TMB molecules by accounting for depth of the potential energy well ($\varepsilon_0 > 0$), and the bonding length of an atomic pair $\sigma$, in between TMB-TMB (H–H, H–C, C–C), TMB-$C_{60}$ (H–$C_{ful}$, C–$C_{ful}$), and $C_{60}$-$C_{60}$ ($C_{ful}$–$C_{ful}$).[39,40] The $C_{60}$-$C_{60}$ and $C_{60}$-TMB interactions are of pure van der Waals nature, while the TMB molecules interact with each other via both van der Waals and Coulomb potentials. The parameters used for calculating the energy of the system are summarized in Table 1.

Note, that it is not a trivial problem to determine the parameters of van der Waals interaction in a multi atomic system. A common way of accounting for the van der Waals interaction is based on the addition of phenomenological Lennard-Jones-type of terms to the total energy of the system[41]. Each of the potentials includes at least two parameters: the equilibrium separation distance of a pair of atoms, and the potential energy well depth. Unfortunately, there are no fixed values for these parameters which would be universally applicable in a wide scope of situations. Even for the same systems different authors choose different parameters.[41-43]

To the best of our knowledge, the van der Waals interaction for the TMB molecules has not been carefully studied so far. There is no experimental information available on noncovalent bonding of two TMB molecules, which is necessary to determine the parameters of the van der Waals interaction between the molecules. Therefore, in the present paper we use the parameters suggested for polycyclic aromatic hydrocarbons,[41] and successfully employed for the study of coronene and circumcoronene clusters.[44] We use this particular set of parameters because it has been shown in Ref. *41* that these parameters can be applied to a large variety of organic molecules including benzene, naphtalene, coronene and others. We assume that parameters suggested in Ref. *41* are reasonable for modelling the non-covalent interaction between the TMB molecules because of the many common features as other studied aromatic molecules. For the van der Waals interaction between the $C_{60}$ molecules we employed the parameters from Ref. *40*, which were derived specially to describe the dispersion forces between fullerenes.

We note that since the van der Waals interaction between the molecules is strongly parameter dependent, the results of this work should be considered as qualitative rather than quantitative. However the relative energies (e.g. the differences in adhesion energy of molecules at different crystallographic planes of the nanowire) should be of the correct order of magnitude, and therefore the present calculation still gives an important insight into the problems of nanowire growth.

We have performed *ab initio* density functional theory calculations to determine the partial charge, $q$, of each atom in the TMB molecule using the B3LYP density functional method as described in the previous section. Since the atoms are bound via covalent polar bonds, all atoms possess partial charges and interact each other via Coulomb potential. The calculated charge distribution of TMB and its optimized structure are shown in Figure 4. It is seen that the charge is significantly different (from −0.40 to +0.32 a.u.) for different carbon atoms in the TMB molecules but it is similar for all the hydrogen atoms (around ~0.1 unit). In a $C_{60}$ molecule, all the C atoms are assumed neutral because of the covalent nonpolar bonds $C_{60}$.

Note that the idea of using the Lennard-Jones type of van der Waals interaction combined with the atomic charges calculated within the framework of the DFT is not new, and was used earlier in the study of coronene clusters,[44] where the bonding between the molecules is also determined by the Coulomb and van der Waals forces.

Having determined the intermolecular potentials and the parameters $\varepsilon_0$, $\sigma$, and $q$, we have calculated the total energy of the system, $U$, using Eq. (13) developed in this work.

$$U = \sum_{\substack{\alpha,\beta=1 \\ \alpha<\beta}}^{N_{C60}} \sum_{\substack{i\in\alpha \\ j\in\beta}} \varepsilon_0 \left[\left(\frac{\sigma}{r_{ij}}\right)^{12} - 2\left(\frac{\sigma}{r_{ij}}\right)^{6}\right] + \sum_{\substack{\alpha,\beta=1 \\ \alpha<\beta}}^{N_{TMB}} \sum_{\substack{I\in\alpha \\ J\in\beta}} \varepsilon_0^{(IJ)} \left[\left(\frac{\sigma^{(IJ)}}{r_{IJ}}\right)^{12} - 2\left(\frac{\sigma^{(IJ)}}{r_{IJ}}\right)^{6}\right] +$$
$$+ \sum_{\substack{\alpha,\beta=1 \\ \alpha<\beta}}^{N_{TMB}} \sum_{\substack{I\in\alpha \\ J\in\beta}} \frac{q_I q_J}{\varepsilon r_{IJ}} + \sum_{\alpha=1}^{N_{TMB}} \sum_{\beta=1}^{N_{C60}} \sum_{\substack{I\in\alpha \\ j\in\beta}} \varepsilon_0^{(Ij)} \left[\left(\frac{\sigma^{(Ij)}}{r_{Ij}}\right)^{12} - 2\left(\frac{\sigma^{(Ij)}}{r_{Ij}}\right)^{6}\right],$$
(13)

where $N_{C60}$ is the total number of the $C_{60}$ molecules in the system, $N_{TMB}$ is the total number of TMB molecules; $\alpha$ and $\beta$ are the fullerene and TMB indices in the first and the second (third) summations correspondingly. In the last term $\alpha$ and $\beta$ numerate TMB and $C_{60}$ molecules, respectively. $i$, $j$ denote atoms inside a fullerene while $I$, $J$ denote atoms inside a TMB molecule. $\varepsilon_0$ and $\sigma$ are parameters of the van der Waals interaction describing the interaction between carbon atoms from two different fullerenes. $\varepsilon_0^{(IJ)}$ and $\sigma^{(IJ)}$ are parameters of van der Waals interaction arising between atoms of two different TMB molecules. Since a TMB molecule consists of hydrogen and carbon atoms, there are three different types of van der Waals interaction: H-H, C-C and C-H. $\varepsilon_0^{(Ij)}$ and $\sigma^{(Ij)}$ are parameters of the $C_{60}$-TMB interaction.

In Eq. (13), $\varepsilon$ is the dielectric constant which reflects the degree of charge screening in the system. In our computation we have assumed $\varepsilon = 1$, which corresponds to the screening-free case. In reality the $\varepsilon$ is larger than 1, and the Coulomb interaction in the system is weaker. However, our calculation shows that the Coulomb energy has a minor influence on the total energy of the system. In the following discussion we demonstrate that the adhesion energy of a unit cell is ~1.0 eV if all Coulomb interactions in the system are accounted for. Without the Coulomb interactions, the adhesion energy of a unit cell changes by only ~1.0 meV, indicating a negligible influence.

By fixing the fullerene molecules to the experimentally observed positions and introducing TMB molecules into the unit cell, we performed structural optimization using conjugate gradient method implemented with the MBN Explorer program.[35] The structures of the stable, low-energy isomeric configurations of the unit cell are shown in Figure 5. The energies calculated for these isomers are indicated in the figure. A specific structure of the unit cell isomer depends on the relative orientation of $C_{60}$ and TMB molecules, and also on the location of the TMB molecules inside the cell. We found that the orientation of the $C_{60}$ molecules may affect the energetic of the unit cell, but has a minor impact on the relative adhesion energy along the three principal growth directions. However, the total energy of the unit cell and the relative adhesion energy largely depends on the location and orientation of TMB molecules inside the cell, for example, in isomer 6, E = -1.528 eV; but in isomer 1, E = -1.899 eV. This relation can be clearly seen in Figure 5.

Figure 6 shows a schematic diagram of the orthorhombic lattice which has three principal growth directions, $a$, $b$, and $c$. We used the adhesion energy as a measure to assess the relative easiness of a unit cell to extend along a specific direction. The smaller the adhesion energy is, the more favorable the crystal grows in the corresponding direction. The adhesion energy of a quadratic monolayer is defined as:

$$E_i = U_N^{(i)} - U_{N-1}^{(i)} - U_1^{(i)} \tag{14}$$

where $U_N^{(i)}$ is the total energy of the crystal with $N$ layers of unit cells along the growth direction, which has been calculated according to Eq. (13); $U_{N-1}^{(i)}$ is the total energy of the same crystal with $N-1$ layers; and $U_1^{(i)}$ is the energy of a single monolayer ($i = a, b, c$). Another important quantity is the adhesion energy of a single unit cell. It is defined as:

$$\varepsilon_i = \frac{E_i}{N_{surf}} = \frac{U_N^{(i)} - U_{N-1}^{(i)} - U_1^{(i)}}{N_{surf}} \tag{15}$$

Here $N_{surf}$ is the number of unit cells on the surface of a nanowire in the direction perpendicular to the growth direction. $\varepsilon_i$ defines the average energy needed to remove a unit cell from the crystal surface. In the calculation we studied the adhesion of 3×3, 4×4 and 5×5 monolayers. For the 3×3 monolayer, we considered a total of 8 layers that are sequentially added one on another. The energies of all intermediate structures were calculated and analyzed. For the 4×4 and 5×5 monolayer, we considered a total of 7 and 4 layers, respectively, and the biggest structure contains over 18,000 atoms.

Figures 7 and 8 show the calculated adhesion energies of a unit cell as a function of the number of the crystalline layers along different growth directions. We found that for most of the isomers, the $b$ growth axis is energetically favorable (adhesion energy along this axis is the smallest). This is the case for all isomers except isomer 1 (Figure 5), for which the adhesion energy along the axis $a$ is the lowest (Figure 8). In addition, the number of the add-on layers has a minor influence on the adhesion energy. For example, the adhesion energy per unit cell in the 3×3 monolayer is only 5 % higher than that in the 4×4 monolayer, while this small difference is even further reduced to ≤ 2 % for between the 4×4 and 5×5 monolayers, indicating that the adhesion energy of a unit cell saturates rapidly.

Having established all the above conceptual items and their quantitative values, we have next considered how the different isomeric structure of the unit cell would affect the crystal shape in the $C_{60}$·TMB nanowires. Figure 9 shows the saturated adhesion energy, $\varepsilon_i^{(0)}$, ($i = a, b, c$), corresponding to the adhesion energy of a unit cell to an infinitely thick crystal having 3×3 and 4×4 unit cells in the cross section. Compared to $\varepsilon_a^{(0)}$ and $\varepsilon_b^{(0)}$, the value of $\varepsilon_c^{(0)}$ is always the highest, suggesting that the growth along $c$ axis would be most difficult, consistent with the experimental observation.

We attribute this restriction to the alternative packing of $C_{60}$ and TMB molecules along the $c$ axis. As a consequence of this, the growth in this direction is substantially limited and the wing thickness is controlled by the size of the crystal core. However, depending on the type of an isomer, the lowest adhesion energy could be along either *a* or *b* direction. The adhesion energy difference between the *a* and *b*, $\Delta = \varepsilon_a^{(0)} - \varepsilon_b^{(0)}$, can be sufficiently large, as for example, $\Delta = -0.30$ eV for isomer 1 and $+0.15$ eV for isomer 6. Such a large energy difference strongly suggests that the crystals with a given isomeric structure would preferably grow along one direction which leads to the formation of a wire-like 1D shape. For isomer 1, this direction is the axis *a*; but for other isomers, it is the axis *b*.

Close examination of the two growth possibilities along either the direction *a* or *b* also indicates that the degree or extent of their growth is different. When certain crystals grow along *a* (such as for isomer 1), their simultaneous growth along *b* would be significantly suppressed because of the much higher adhesion energy in *b* direction (~0.30 eV higher). In this case, the crystals favourably grow almost only along *a*, which could result in the development of a thin nanowire with narrow wings or even wingless. However, if a crystal grows along *b* (such as for isomer 3), its simultaneous growth along *a* may still be able to reasonably develop because of the relatively smaller adhesion energy difference between the *a* and *b* ($\Delta_{a,b} = 0.07 - 0.15$ eV, depending on the structure of the isomer). This is the case where a crystal is more likely to grow into a thick nanowire with wider wings. Because there are more isomeric structures favouring the growth along *b* axis, as illustrated in Figures 5 to 9, it is expected that in comparison with the thin nanowires, more crystals would grow into thick ones possessing wider wings, and this has been confirmed by our extensive electron microscopic observations.

## 5. Conclusion

We have investigated the growth of fullerene-based nanowires ($C_{60}$·TMB) which exhibit a highly interesting but unusual morphology featured by an overall Y-shaped crystal cross section. Nucleation of such nanowires was found to start from the central core area, and the structure had a significant orthorhombic distortion during the growing process because of the preferential introduction of the solvent molecules into the crystal lattice along the principal axes. We also show that depending on location and orientation of the solvent molecules in the crystal unit cell, the adhesion energy along the principal growth direction can be 0.1-0.3 eV lower than that along other directions, offering an explanation why the nanocrystals tend to grow into a 1D shape. Because our treatment to the system follows the universal van der Waals and Coulomb potentials, we propose that the approach demonstrated in this work may be widely useful to explore other host-guest interactions in crystal engineering in an attempt to tune the crystal shape and structure.

**Acknowledgement.** This work was supported by the European EXCELL and CANAPE project. The possibility to perform complex computer simulations at the Frankfurt Centre for Scientific Computing is also gratefully acknowledged.

**Table 1.** Parameters of the van der Waals interactions used for calculating the energy of the system. The data for the TMB-TMB interaction (H-H, C-C, H-H) are taken from Ref. *39*. The parameters for the interaction of hydrogen atom from a TMB molecule with a carbon atom from a $C_{60}$ (H-$C_{ful}$) are also adapted from Ref. *39*. The parameters describing the interaction of carbon atoms from two $C_{60}$ molecules ($C_{ful}$-$C_{ful}$) and for the interaction of a carbon atom from a TMB molecule with a carbon atom from a $C_{60}$ (C-$C_{ful}$) are adapted from Ref. *40*.

|  | *H-H* | *H-C* | *C-C* | *H-$C_{ful}$* | *$C_{ful}$-$C_{ful}$* | *C-$C_{ful}$* |
|---|---|---|---|---|---|---|
| $\varepsilon_0$ (meV) | 0.563 | 1.487 | 4.069 | 1.487 | 2.860 | 2.620 |
| $\sigma$ (Å) | 3.297 | 3.601 | 3.901 | 3.601 | 3.890 | 3.860 |

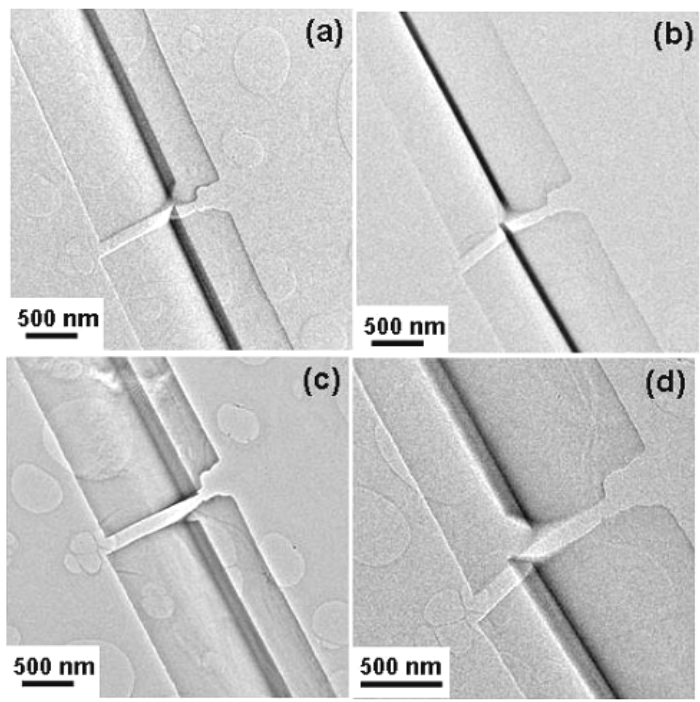

**Figure 1.** A set of TEM images of a thick but broken nanowire recorded with varied tilting angles of the sample grid (from a to d) to show the overall morphology and the three nanobelt-like wings.

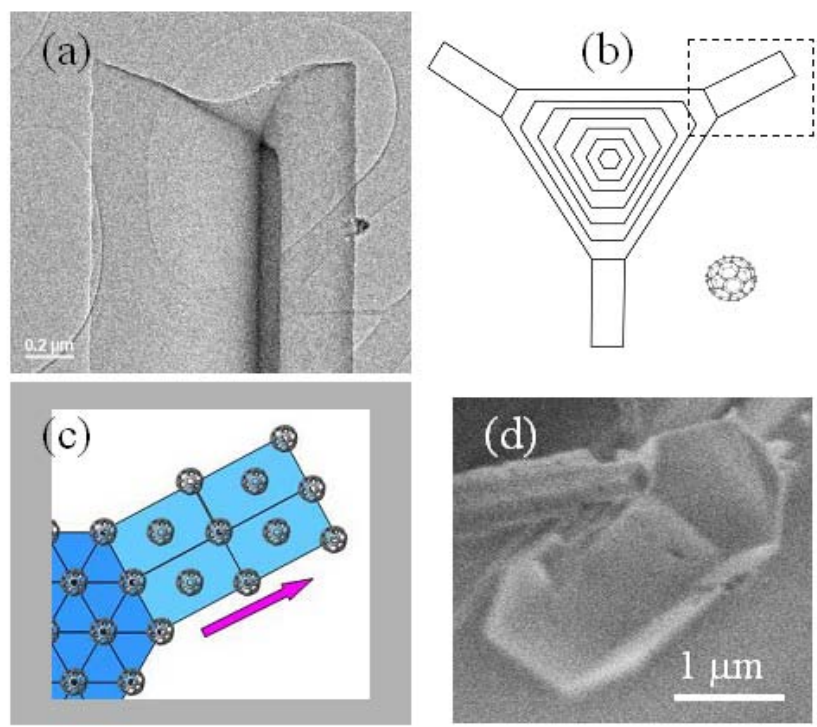

**Figure 2.** (a) A TEM image of a $C_{60}$·TMB nanowire shows the cross section of the crystal. (b) A schematic drawing to indicate that the cross section may be viewed as being a triangle shape which starts from a hexagonal core but the shape gradually transformed in the subsequent growth process due to inclusion of the solvent molecules into the lattice. (c) An enlarged schematic diagram, corresponding to the square area marked in (b), to show the $C_{60}$ arrangement at the interface of the core and the wing. (d) A SEM image of a by-product nanocrystal with a hexagonal shape observed in the same batch of sample.

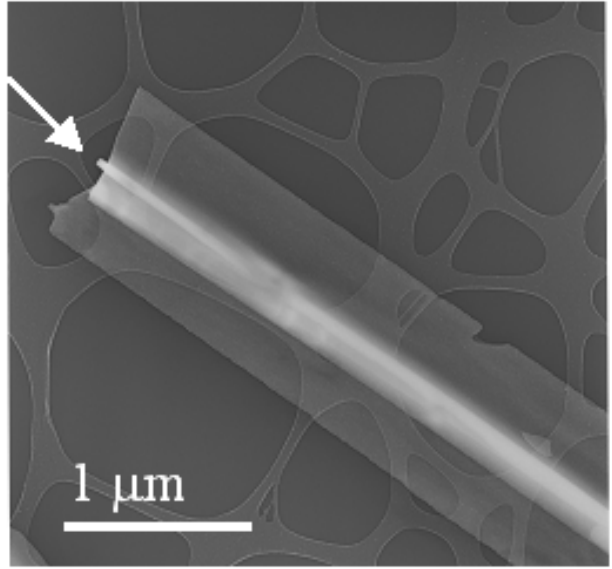

**Figure 3.** A TEM image (negative film) shows the 'over-grown' tip as an extension of the crystalline core.

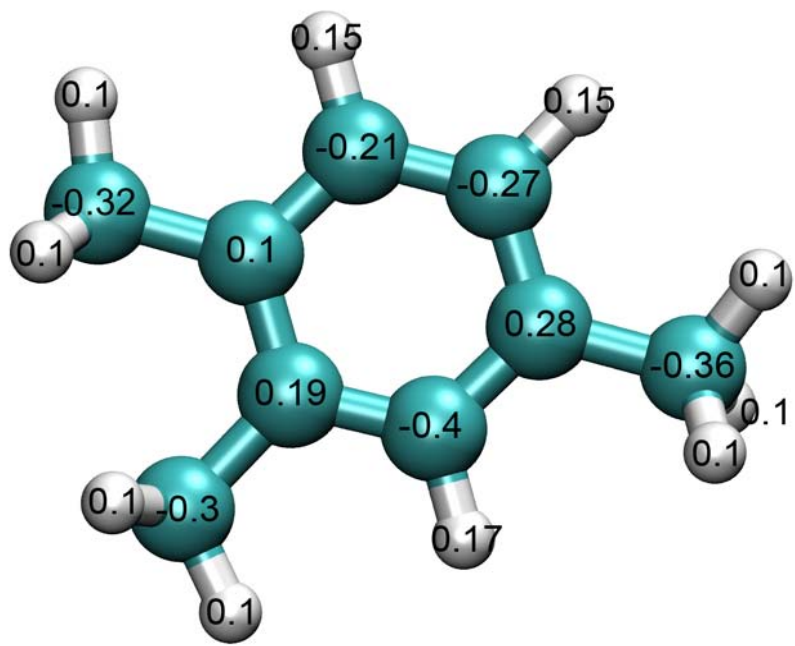

**Figure 4.** Optimized structure of a 1,2,4-TMB molecule and the corresponding charge distribution calculated using the B3LYP method combined with the standard 6-31G(d) basis set.

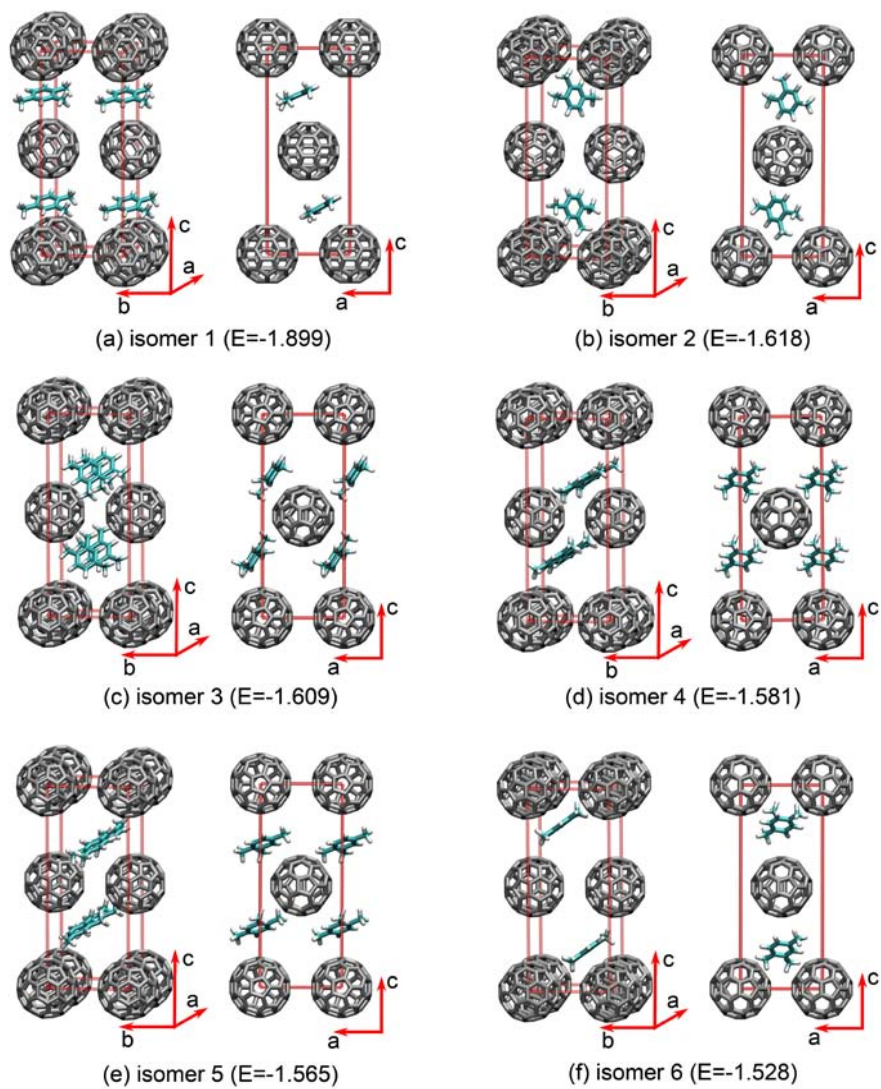

**Figure 5.** Optimized isomeric states in the $C_{60}\cdot$TMB nanowire unit cell as derived from the calculations. The number in the brackets below each image shows the energy of the structure (in eV). The coordinate frames used in the present work are also indicated.

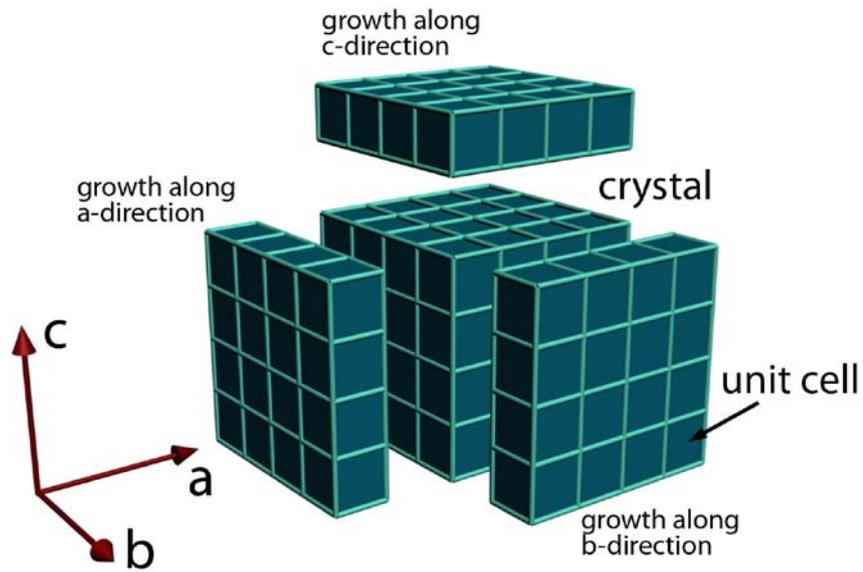

**Figure 6.** A schematic diagram to show an orthorhombic crystal that can grow in three principal directions. To understand a crystal growth along different directions, we investigated the adhesion of crystalline layers to the crystal. Here we illustrate how a monolayer, with 4×4 unit cells, as an example, can be attached to the three principal faces of the crystal.

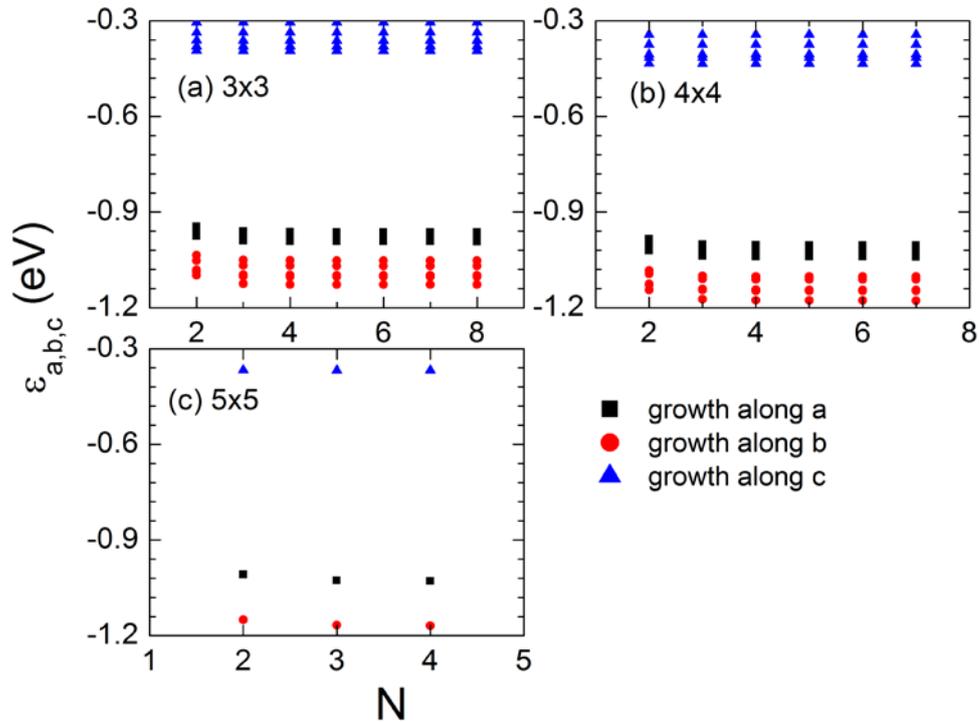

**Figure 7.** Adhesion energy of a cell calculated as a function of the number of crystalline layers for different crystal growth directions. The result corresponds to isomers 'b' to 'f', shown in Figure 5. The adhesion energy for growth along *a*-, *b*- and *c*-direction of the crystal is shown with squares, circles, and triangles, respectively. The plots (a), (b) and (c) correspond to adhesion of a monolayer consisting of 3×3, 4×4 and 5×5 cells, respectively.

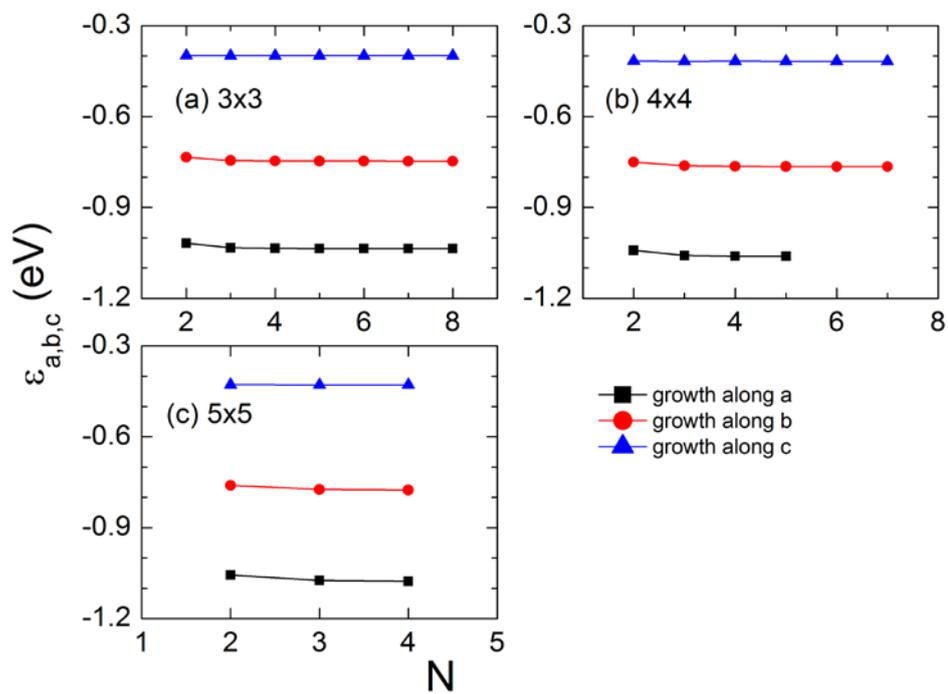

**Figure 8.** The same calculation as that in Figure 7, but performed for the isomer 'a' as shown in Figure 5.

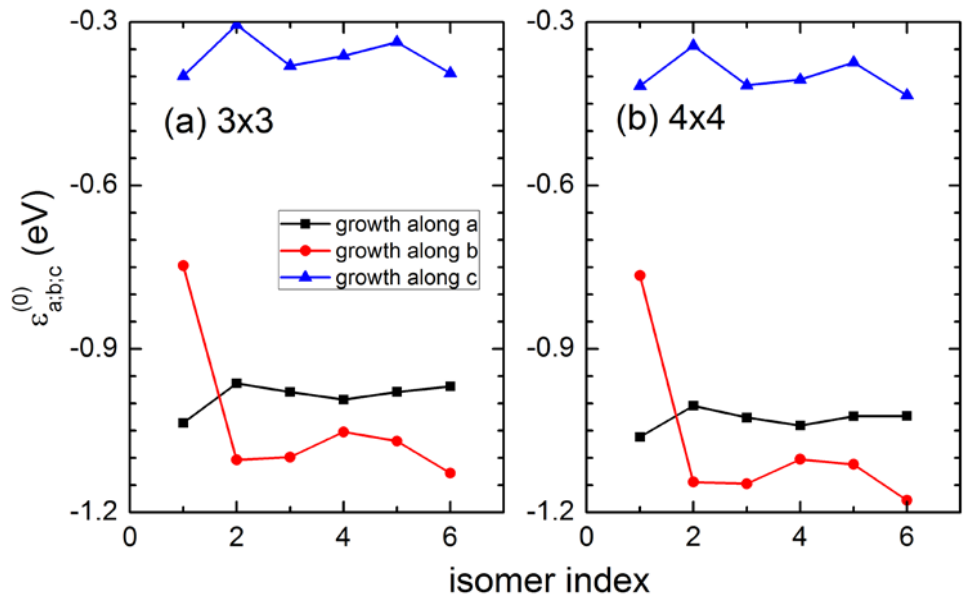

**Figure 9.** Adhesion energy of a unit cell to an infinitely thick crystal along the *a*- (squares), *b*- (circles) and *c*- nanowire growth directions (triangles) calculated for different isomeric states of the crystal's unit cell. Plots (a) and (b) correspond to the adhesion of a nanolayer consisting of 3×3 and 4×4 cells respectively.